\documentclass[prl,aps,showpacs,floatfix,floats,superscriptaddress,nobalancelastpage,twocolumn]{revtex4-1}
\usepackage{graphicx}
\usepackage{graphics}
\usepackage{latexsym,amsmath,amssymb,bm,euscript}
\usepackage{color}
\usepackage{dcolumn}
\usepackage{bm}
\usepackage{epstopdf}
\usepackage{times}
\usepackage{multirow}
\usepackage{wasysym}
\usepackage{subfigure}
\usepackage{bbm}
\begin{document}

\title{Probing Critical Surfaces in Momentum Space Using Real-Space Entanglement Entropy:\\ Bose versus Fermi}
\author{Hsin-Hua Lai}
\affiliation{Department of Physics and Astronomy, Rice University, Houston, Texas 77005, USA}
\affiliation{National High Magnetic Field Laboratory, Florida State University, Tallahassee, Florida 32310, USA}
\author{Kun Yang}
\affiliation{Department of Physics and National High Magnetic Field Laboratory, Florida State University, Tallahassee, Florida 32306, USA}
\date{\today}

%%%%%%%%%%%%%%%%%%%%%%%%%%%%%%%%%%%%%%%%%%%%%%%%%%%%%%
\begin{abstract}
A co-dimension one critical surface in the momentum space can be either a familiar Fermi surface, which separates occupied states from empty ones in the non-interacting fermion case, or a novel Bose surface, where gapless bosonic excitations are anchored. Their presence gives rise to logarithmic violation of entanglement entropy area law. When they are \textit{convex}, we show that the shape of these critical surfaces can be determined by inspecting the leading logarithmic term of real space entanglement entropy. The fundamental difference between a Fermi surface and a Bose surface is revealed by the fact that the logarithmic terms in entanglement entropies differ by a factor of two: $S^{Bose}_{log} = 2 S^{Fermi}_{log}$, even when they have identical geometry. Our method has remarkable similarity with determining Fermi surface shape using quantum oscillation. We also discuss possible probes of \textit{concave} critical surfaces in momentum space.
\end{abstract}
\maketitle

%%%%%%%%%%%%%%%%%%%%%%%%%%%%%%%%%%%%%%%%
\textit{Introduction} --
Various aspects of quantum entanglement \cite{QE_RMP} have been extensively studied in recent years. The most widely used measure of entanglement is the entanglement entropy (EE), which is the von Neumann entropy associated with the reduced density matrix of a subsystem, obtained by tracing out degrees of freedom outside it. For extended quantum systems, it is generally believed that ground states of all gapped local Hamiltonians, as well as a large number of gapless systems, follow the so-called area law, which states that the EE is proportional to the surface area of the subsystem \cite{Eisert_RMP}. Violations of the area law, usually in a logarithmic fashion, do exist in various systems. In one dimension (1D), they are found to be associated with quantum criticality \cite{Calabrese2004,Refael2004,Laflorencie2005, Bonesteel2007, Bravyi2012}. Above 1D such violations are very rare. The well-established examples are systems with \textit{Fermi surfaces}, including free fermion ground states \cite{Wolf,GioevKlich,Swingle}, and Fermi liquid phases \cite{dingprx12}. For the case of \textit{Bose surfaces}, defined as co-dimension one surfaces in momentum space where gapless {\em bosonic} excitations live, we constructed harmonic lattice models with short-range couplings and found a similar area-law violation\cite{Lai_EE4EBL}, which realized a lattice version of the Exciton Bose Liquid (EBL) phase \cite{Paramekanti02, Tay10}. Strongly-interacting systems with {\em emergent} Fermi surfaces have also been studied numerically with evidence of area-law violation as well \cite{Zhang11, EE_HLR}. In the present paper we explicitly focus on the systems above 1D and refer to Fermi and Bose surfaces jointly as critical surfaces (in momentum space).

It is impossible to overstate the importance of such critical surfaces to the long-distance/low-energy physics of the system. However unlike the free fermion/harmonic oscillator systems where their presence and shapes are ``obvious", in strongly-interacting systems they may be associated with heavily renormalized degrees of freedom or emergent, and thus difficult to detect (either theoretically from the Hamiltonian or ground state wave function, or experimentally). Recently it was suggested that logarithmic violation of entanglement entropy area law is an effective, and sometimes unique way to probe the presence of Fermi surfaces in strongly interacting systems\cite{RyuTakayanagi2006, Tatsuma2009, Zhang11, Ogawa2012, Huijse12, EE_HLR}.

The purpose of this paper is three-fold. First of all, as already mentioned above, Fermi and Bose surfaces {\em both} give rise to logarithmic violation of entanglement entropy area law; thus such violation indicates the presence of critical surface(s), but not necessarily Fermi surface(s). Due to the similar effect they have on EE, one might think they are equivalent. We reveal their qualitative difference by demonstrating the presence of a factor of two difference in their contribution to the logarithmic term in EE. We further demonstrate that not only the presence, but also the shape of such critical surface can be determined from inspecting the scaling behavior of EE. This is particularly true when these surfaces are convex, in which case our (theoretical) method has remarkable similarity with determining a Fermi surface shape experimentally using quantum oscillation \cite{Book-Shoenberg}. Lastly we argue that with some additional input, we may be able to distinguish between Bose and Fermi surfaces.

\begin{figure}[t]
   \centering
   \includegraphics[width=3 in]{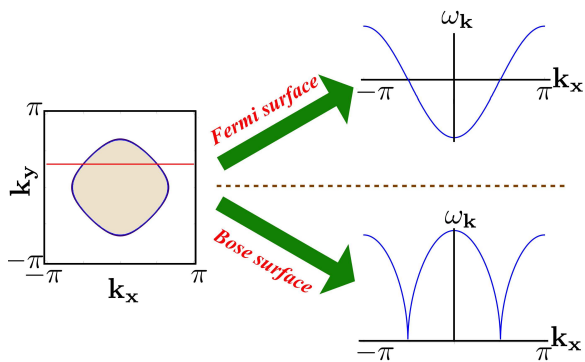}
   \caption{(Color Online) A Fermi surface versus a Bose surface. Left panel shows a critical (zero energy) surface represented by the closed blue line in momentum space. The sharp difference between a Bose and a Fermi surface can be revealed by inspecting the dispersion along a line that cuts across the surface represented by the red line. The top right (bottom left) panel shows the case of a Fermi surface (Bose surface), where the blue line represents an 1D fermionic (bosonic) dispersion, with one (two) pair(s) of left and right moving modes crossing zero energy.
 }
   \label{Fig:Bose_vs_Fermi}
\end{figure}

%%%%%%%%%%%%%%%%%%%%%%%%%%%%%%%%%%%%%%%%%%%%
\textit{Critical surfaces: Bose v.s. Fermi}--
The left panel of Fig.~\ref{Fig:Bose_vs_Fermi} shows an extensive critical surface in momentum space. The subtle difference between a Bose surface and a Fermi surface is best revealed by inspecting the dispersion along a line that cuts through the surface (as illustrated by the red line). The corresponding dispersions are shown on the right panels of Fig.~\ref{Fig:Bose_vs_Fermi}. The top right panel shows the usual 1D fermion dispersion with a pair of left and right moving Fermi points crossing the Fermi energy. The bottom right panel illustrates the 1D gapless Boson dispersion which touches the zero energy twice.
It should be clear that the low-energy modes at one specific intersection point are chiral when it is of the Fermi type, while they are non-chiral for the Bose-type intersection. This leads to the factor of two difference in their contribution to EE mentioned above, as we now elaborate. Before proceeding, we emphasize that the critical surface discussed in the present work refers to the surface formed in the momentum space by the gapless fermionic or bosonic degrees of freedom ``emergent" at the long-wavelength (low energy) scale in an interacting model system, instead of the trivially noninteracting cases. Furthermore, in the noninteracting limit, there should be \textit{no} (extended) Bose surface since for a typical critical bosonic system the gapless bosonic degree of freedom only live at a single gapless point instead of living in a extended Bose surface, which will not lead to a leading area-law violated entanglement entropy \cite{cramer2007statistics}. 

If we had 1D systems with the fermionic/bosonic dispersions of the top/bottom panels of Fig.~\ref{Fig:Bose_vs_Fermi}, they would correspond to conformal field theories (CFTs) with central charges $c=1$ and $c=2$ respectively, and EE would scale with subsystem size as $S_F^{1D} \dot{=} \frac{1}{3} \ln L$ for Fermi and $S^{1D}_B \dot{=} \frac{2}{3} \ln L = 2S^{1D}_F$ for Bose, where $L$ is the subsystem length and $\dot{=}$ represents the leading contribution of the EE. In higher dimensions the logarithmic enhancement of EE can be understood by dividing the critical surfaces into patches small enough so that within each patch their curvatures may be neglected, and map them onto effective 1D systems described by CFTs, and sum up their contributions to EE \cite{Swingle,dingprx12}. We thus see that for Bose and Fermi surfaces with identical shape, their contribution to the leading term in EE differ by a factor of two, revealing their qualitative difference.

For probing a critical surface in momentum space in $d$ dimensions, we first present the general formula for the leading logarithmic term in EE. In $d$ dimensions, we consider a specific real-space partition in which the boundary between the two subsystems is a plane whose normal direction is $\hat{n}_d$. This partition preserves the translational symmetries in $d-1$ dimensions that are perpendicular to $\hat{n}_d$, and we follow the similar procedures used in Refs.~\cite{cramer2007statistics,Lai_EE4EBL} to perform partial Fourier transformation for all the physical degrees of freedom along these $d-1$ axes, since the momenta $k_{1,2,\cdots, n-1}$ are good quantum numbers. We thus view the momentum space as consisting of arrays of parallel 1D chains with spacings $\delta k_{1,2,...,n-1} = 2\pi/L_\bot$, where $L_\bot$ is the linear size of these transverse directions.

As stated above, each 1D line intersecting the critical surface contributes $(\xi_a/3) \ln \mathcal{L}_\|$ to the leading term of the EE, where $\xi_{a = F, B} $ with $\xi_F =1 (\xi_B = 2)$ for a Fermi (Bose) surface, and $\mathcal{L}_\|$ is the linear size of the (smaller) subsystem along $\hat{n}_d$. The total leading EE can be obtained by counting total number of chains (in momentum space) intersecting the critical surface, which is the cross-sectional area of the critical surface divided by the $(d-1)$ dimensional spacing area between each chain, $(2\pi/L_\bot)^{d-1}$. Explicitly, the leading term of the EE is
\begin{eqnarray}
\nonumber S_{dD} &\dot{=}& \frac{\xi_a}{3} \ln \mathcal{L}_\| \times \frac{1}{2} \times \frac{\int_{\partial \Gamma} \left| d\hat{S}_\Gamma \cdot \hat{n}_d\right|}{(2\pi/L_\bot)^{d-1}} \\
\nonumber & = & \frac{\xi_a}{3}\ln \mathcal{L}_\| \times \frac{1}{2} \times \left( \frac{L_\bot}{2\pi}\right)^{d-1} \times \frac{\int_{\partial \Gamma}\int_{\partial A} \left | d\hat{S}_\Gamma \cdot d\vec{S}_A\right|}{2L_\bot^{d-1}} \\
& = & \frac{\xi_a}{12} \frac{\ln \mathcal{L}_\|}{(2\pi)^{d-1}} \int_{\partial A} \int_{\partial \Gamma} \left| d\vec{S}_A \cdot d\hat{S}_\Gamma \right|,\label{Eq:WGK}
\end{eqnarray}
where the factor of $1/2$ at the first line is due to the over counting of the cross-section. In second line, we rewrite $\hat{n}_d$ as real-space partition surface integral (with $d\vec{S}_A$ being the corresponding oriented area element whose direction is along the local normal direction) divided by the partition surface area in $d-1$ dimensions, $2L_\bot^{d-1}$. $\int_{\partial \Gamma}$ represents the surface integral along the critical surface in momentum space (with $d\hat{S}_\Gamma$ being the corresponding oriented area element). While we arrived at Eq.~(\ref{Eq:WGK}) by considering the special partition we will use later, it is actually the correct formula for free fermion state for arbitrary cuts \cite{GioevKlich}, if we replace $\mathcal{L}_\|$ by the generic linear size of the smaller subsystem. Using arguments along those of Refs. \cite{Swingle,dingprx12}, we conclude that it apply to systems with Bose surfaces with arbitrary partition as well, which is a {\em new} result.

\textit{EE probing a convex critical surface}--
We discuss how to reconstruct the critical surface using EE, based on Eq.~(\ref{Eq:WGK}). The key point for the construction is that the prefactor of $(\xi_a/3) \ln \mathcal{L}_\|$ in Eq.~(\ref{Eq:WGK}) gives the critical surface's cross-sectional area along $\hat{n}_d$,  when it is \textit{convex}.
We leave the discussions on a \textit{concave} critical surface toward the end of the paper.

\begin{figure}[t]
   \centering
   \includegraphics[width=\columnwidth]{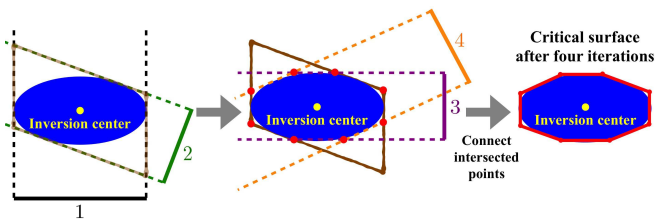}
   \caption{(Color Online) Illustration of reconstructing a 2D critical surface with an inversion center. (Left panel) We start from the largest cross-section line $1$ and rotate the partition to obtain cross-section $2$. They form a parallelogram whose center is the inversion center. (Middle panel) We again rotate the partition to extract $3$, whose dashed lines should be of equal distance to the inversion center. We can keep rotating the partition to extract $4$, $5$, and etc. and appropriately place them around the inversion center. (Right panel) This leads to an approximation of the critical surface by a polygon, with arbitrary accuracy.}
   \label{Fig:geo_map_surface}
\end{figure}

Figure~\ref{Fig:geo_map_surface} illustrates how we reconstruct a convex critical surface using EE in 2D. We note that in this \textit{Letter} we focus on a critical surface in momentum space \textit{with} an inversion center. \cite{Inversion_center}  We start from a direction that gives the largest cross-section represented by the black line $1$. Now rotating the partition direction by an angle $\theta$ to extract the second projected length (green line) $2$. In the real situation, the only information we will have are the angles $\theta$ and the projected lengths at different angles $\ell(\theta)$, while the critical surface is an abstract object that can not be seen (The critical surface shown in Fig.~\ref{Fig:geo_map_surface} is for presentation purpose only).  We now have projected lines $1$ and $2$, but in order to have a reference point for mapping out the convex critical surface shape, we first fix the location of $1$ and arbitrarily place $2$ as long as the dashed lines of $1$ and $2$ intersect (which is always possible since the dashed lines are infinitely long). The intersections between dashed lines $1$ and 2 forms a parallelogram shown in the left panel of Fig.~\ref{Fig:geo_map_surface}, whose center is also the inversion center. At this point the critical surface is approximated by this parallelogram (choice of the location of the inversion center is arbitrary, as it is a gauge-dependent quantity). We continue rotating the partition to extract the projected length $3$ (purple line) in the middle panel of Fig.~\ref{Fig:geo_map_surface}. Since there is an inversion center, the dashed line $3$ must be placed in a position of equal distance to the center. We can continue to rotate the partition to extract the project length $4$ and repeat the procedures, but in this illustration we stop at the fourth iteration. Connecting all the intersected points represented by the red dots in middle panel of Fig.~\ref{Fig:geo_map_surface}, we can \textit{geometrically} extract the qualitative shape of the critical surface, right panel of Fig.~\ref{Fig:geo_map_surface}. It is clear by now that the critical surface can be approximated by a polygon in this case, with {\em arbitray} accuracy. \cite{Radon_Transform}

\begin{figure}[t]
   \centering
   \includegraphics[width=\columnwidth]{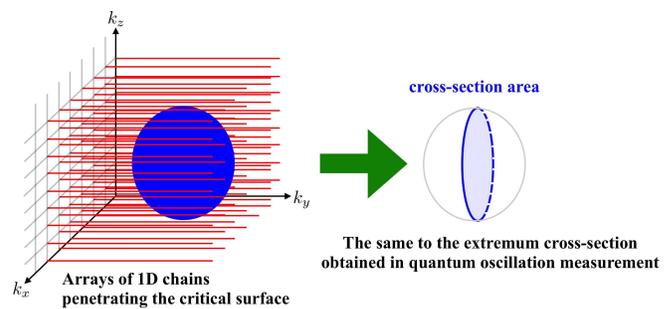}
   \caption{(Color Online) Illustration of extracting the 2D cross-section of a 3D critical surface (blue region). We assume that momenta ${\bf k}_x$ and ${\bf k}_z$ remain good quantum numbers and the momentum space consists of these arrays of 1D chains (red lines). Remarkably, extracting the cross-section using EE in 3D shares similarities with identifying the Fermi surface shape using quantum oscillations. The two approaches give the same cross-section for a \textit{convex} critical surface.}
   \label{Fig:3D_surface}
\end{figure}

In three dimensions (3D), we find that the EE construction of a convex critical surface shape share remarkable similarities with experimentally identifying a Fermi surface shape using quantum oscillation \cite{Book-Shoenberg}. Figure~\ref{Fig:3D_surface} gives an illustration of extracting the cross-section in 3D using Eq. (\ref{Eq:WGK}). The relation between them is the following. Quantum oscillation measurement determines the cross-section areas of {\em all} Fermi surface \textit{extrema} perpendicular to the magnetic field direction, which include both the maximum cross-section and the minimum cross-section. The EE probes the projectional area of the critical surface along any direction (equivalent to magnetic field direction in quantum oscillation experiment), which is {\em equal} to the maximum cross-section. This is the only extremum when the critical surface is convex, and in this case the two methods are {\em identical}. One can thus use the same algorithm in quantum oscillation measurements here to determine the Fermi surface shape \cite{Book-Shoenberg}. 

\textit{Discussion}--
As discussed earlier, the presence of logarithmic enhancement of EE indicates presence of critical surface(s) above 1D. However, if we do not know which type of the critical surface (Fermi or Bose) leading to the logarithmic enhancement, using the wrong version of Eq.~(\ref{Eq:WGK}) results in a numerical error in the size of the critical surface, although one would still get the correct shape. EE itself does not distinguish between Fermi and Bose surfaces. We note when combined with other indicators, one may be able to make a distinction. Fermi surface volumes often obey the Luttinger's theorem \cite{Luttinger1960}, from which one can check if it is consistent with that of the critical surface obtained using the Fermi version of Eq.~(\ref{Eq:WGK}); if not then a possible interpretation is that the logarithmic enhancement of EE originates from Bose surface(s). Another example is a circular/spherical Fermi surface often gives rise to Friedel oscillations in ground state density-density correlation function with wave vector $2{\bf k}_F$ \cite{Friedel}, which can be used to perform a similar check.

\begin{figure}[t]
   \centering
   \includegraphics[width=3 in]{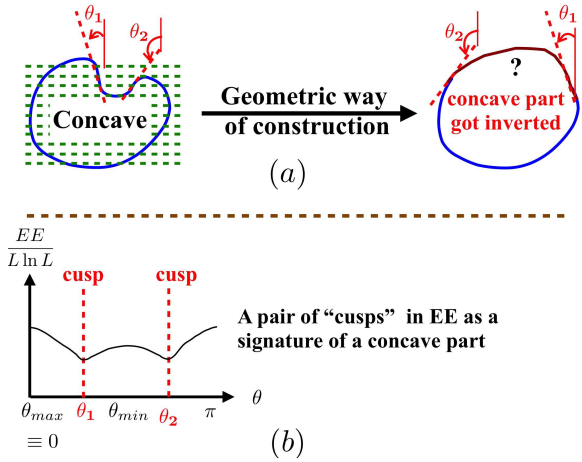}
   \caption{(Color Online) Probing a concave critical surface using entanglement entropy. (a) Applying the method for a convex critical surface may lead to a surface shape with the \textit{inverted} concave part, where we assume the concavity occurs at $\theta \in [\theta_1, \theta_2]$ with $\theta_1, \theta_2 \in [0,\pi]$ due to the $\pi$-period in EE. (b) The cusp-like signatures due to the presence of a concave structure at $\theta \in [\theta_1, \theta_2]$ in the angular dependence of EE, where we assume EE is maximum at the starting angle.}
   \label{Fig:Concave}
\end{figure}

So far we have focused on a \textit{convex} critical surface, but in general a \textit{concave} critical surface is possible. For such a concave surface, the method presented here \textit{cannot} completely determine its shape. However, applying our method can in principle \textit{infer} the location of the concave part; see Fig.~\ref{Fig:Concave} for illustration. Fig.~\ref{Fig:Concave}(a)(top panel) illustrates a typical concave surface, whose concavity occurs between $\theta\in [\theta_1, \theta_2]$ with $\theta_1, \theta_2 \in [ 0, \pi]$ due to $\pi$-period in EE and $\theta_1$, $ \theta_2$ being directions along which there exist tangential lines that go through turning points of the curve. Blindly applying Eq.~(\ref{Eq:WGK}) to this case the same way as the convex case would lead to a wrong shape with its concave part replaced by a convex shape (inverted concave shape). The reason is due to the fact that the EE probe effectively measures the total number of intersected points between the 1D chains consisting of the momentum space and the critical surface. At $\theta \in [\theta_1, \theta_2]$, some 1D chains in the momentum space intersect the surface \textit{four} times, as shown in Fig.~\ref{Fig:Concave}(a), and they contribute additional EE, resulting in the wrong shape. We note that the wrong shape alters the volume (area) of the critical surface and would lead to the violation of the Luttinger's theorem, which can serve as an alarm of the presence of the concavity of the surface.

There are, however, signs that the critical surface is actually concave. To reveal them we suggest that one measures the leading EE, $S_{log}$, as a function of angle $\theta$, as shown in Fig.~\ref{Fig:Concave}(b). For a convex shape, if we begin at an angle $\theta \equiv 0$ where $S_{log}(\theta)$ is the maximum, $S_{log}(\theta)$ should smoothly decrease until it hits the minimum corresponding to the minimum cross-section, after that $S_{log}(\theta)$ smoothly increasing back to the initial value at $\theta = \pi$. For a concavity at $\theta \in [ \theta_1, \theta_2]$, Fig.~\ref{Fig:Concave}, if we start from the maximum $S_{log}(\theta)$ (we assume it occurs at $\theta \not\in [\theta_1, \theta_2]$ for a small concavity), $S_{log}(\theta)$ would smoothly decrease until $\theta_1$, where the additional contributions to $S_{log}$ due to the additional intersected points between the 1D momentum-space chains (green lines in Fig.~\ref{Fig:Concave}) and the critical surface would upward the curve resulting in a \textit{cusp}-like structure at $\theta_1$. \cite{concave_assumption} For $\theta > \theta_1$, $S_{log}(\theta)$ increases until it arrives at a maximum when the total number of intersected points are maximum, and, after that point, it smoothly decreases until $\theta = \theta_2$. For $\theta > \theta_2$, the $S_{log}(\theta)$ picks up the correct convex structure and smoothly increase to the initial maximum EE at $\theta = \pi$, resulting in another cusp structure at $\theta_2$. The pair of cusps can serve as evidence of the presence of a concave part in the critical surface between $\theta \in [\theta_1, \theta_2]$. 

We note that the method presented here uses a specific type of partition (that respects translation symmetry in all transverse directions). It does not fully utilizes the power of Eq. (\ref{Eq:WGK}). It is quite possible that by considering more general types of partitioning we will be able to determine generic shapes of critical surfaces. This will be left for future work.

The EE probes of the critical surfaces in momentum space can be possibly applied to the quantum spin models that realize the gapless spin liquids with fermionic spinon surfaces \cite{LeeandLee05,ringxch, Sheng09, Block2010, Mishmash11,Chua11,Biswas11,Baskaran09,Lai11, Hermanns2014}. The exactly solvable quantum spin models realizing the (Majorana) fermionic surfaces \cite{Chua11, Baskaran09, Lai11, Hermanns2014} can serve as the promising models to test the construction of the critical surfaces using EE since we know exactly the shapes of the critical surfaces. For the more complex theoretical models which realize the spin Bose metal with fermionic spinon Fermi sea coupled to U(1) gauge fields \cite{LeeandLee05, ringxch, Sheng09, Lai10,Mishmash2015},  composite fermion Fermi liquid state \cite{EE_HLR}, and certain theoretical models realizing the non-Fermi liquid metal phases \cite{DBL, Sheng09, Jiang2013}, the EE probes may well be the only method in mapping out the critical surfaces in momentum space. A possible issue regarding these states is the U(1) gauge fluctuations \cite{LeeNagaosaWen}, which may modify Eq.~(\ref{Eq:WGK}) and may be viewed as indication of the breakdown of Fermi liquid behavior. 

More specifically, the effective central charge $\xi_{s}$ associated with the entanglement entropy contributed from each pair of critical surface patches, which can be viewed as an effective 1D system without conformal invariance in the charge sectors due to the presence of gauge fields, is unknown. The direct consequence is that the EE probe can only gives the overall value of the multiplication of the cross-sectional area and the effective central charge, $\xi_{s} A$, after we eliminate the common factor, $(L/3) \ln L$. In order to apply our algorithm to map out such critical surface, we need to determine $\xi_{s}$. Since $\xi_{s}$ is associated with the effective 1D system of the critical surface patches, it does not depend on the size or shape of the critical surface. We can focus on the isotropic case and numerically calculate a relevant correlation function for an arbitrary observation direction, i.e., the spin-spin correlation function for the spin Bose metal phases with spinon Fermi sea. The power-law correlations in the real space correspond to the singularities in the momentum space and the corresponding structure factor will show singular behaviors at wave vector ${\bf q} =0$ and at wave vector ${\bf q} = {\bf k}_{FR} - {\bf k}_{FL}$, where we introduce the wavevector ${\bf k}_{FR} = - {\bf k}_{FL} \equiv {\bf k}_F$ (since it's a circular critical surface), of a right/left patch of the critical surface whose unit surface vector (which is perpendicular to the surface) parallel/antiparallel to the observation direction. Most important of all is that $|{\bf k}_{FR} - {\bf k}_{FL}|$ is exactly the cross-section (diameter) $A$ of the spinon critical surface. Comparison between the result obtained in the correlation function calculations and that obtained in real-space EE calculation can determine $\xi_{s}$.

%%%%%%%%%%%%%%%%%%%%%%%%%%%%%%%%%%%%%%%%%%%
\textit{Conclusion} --
In this paper we have shown how to determine the geometries of critical surfaces in momentum space using real space entanglement entropy of the ground state, and possibly distinguish between Bose and Fermi surfaces, which are qualitatively different.

%%%%%%%%%%%%%%%%%%%%%%%%%%%%%%%%%%%%%%%%%%%%%%%%%%%%%%%%%%
\textit{Acknowledgments}--HHL and KY acknowledge the National Science Foundation through Grants No. DMR-1004545, DMR-1157490, No. DMR-1442366, and State of Florida. HHL is also partially supported by National Science Foundation through Grants No. DMR -1309531, DMR - 1350237, and the Smalley Postdoctoral Fellowship in Quantum Materials at Rice University.
%%%%%%%%%%%%%%%%%%%%%%%%%%%%%%%
\bibliography{biblio4EEprob}

%merlin.mbs apsrev4-1.bst 2010-07-25 4.21a (PWD, AO, DPC) hacked
%Control: key (0)
%Control: author (8) initials jnrlst
%Control: editor formatted (1) identically to author
%Control: production of article title (-1) disabled
%Control: page (0) single
%Control: year (1) truncated
%Control: production of eprint (0) enabled
\begin{thebibliography}{43}%
\makeatletter
\providecommand \@ifxundefined [1]{%
 \@ifx{#1\undefined}
}%
\providecommand \@ifnum [1]{%
 \ifnum #1\expandafter \@firstoftwo
 \else \expandafter \@secondoftwo
 \fi
}%
\providecommand \@ifx [1]{%
 \ifx #1\expandafter \@firstoftwo
 \else \expandafter \@secondoftwo
 \fi
}%
\providecommand \natexlab [1]{#1}%
\providecommand \enquote  [1]{``#1''}%
\providecommand \bibnamefont  [1]{#1}%
\providecommand \bibfnamefont [1]{#1}%
\providecommand \citenamefont [1]{#1}%
\providecommand \href@noop [0]{\@secondoftwo}%
\providecommand \href [0]{\begingroup \@sanitize@url \@href}%
\providecommand \@href[1]{\@@startlink{#1}\@@href}%
\providecommand \@@href[1]{\endgroup#1\@@endlink}%
\providecommand \@sanitize@url [0]{\catcode `\\12\catcode `\$12\catcode
  `\&12\catcode `\#12\catcode `\^12\catcode `\_12\catcode `\%12\relax}%
\providecommand \@@startlink[1]{}%
\providecommand \@@endlink[0]{}%
\providecommand \url  [0]{\begingroup\@sanitize@url \@url }%
\providecommand \@url [1]{\endgroup\@href {#1}{\urlprefix }}%
\providecommand \urlprefix  [0]{URL }%
\providecommand \Eprint [0]{\href }%
\providecommand \doibase [0]{http://dx.doi.org/}%
\providecommand \selectlanguage [0]{\@gobble}%
\providecommand \bibinfo  [0]{\@secondoftwo}%
\providecommand \bibfield  [0]{\@secondoftwo}%
\providecommand \translation [1]{[#1]}%
\providecommand \BibitemOpen [0]{}%
\providecommand \bibitemStop [0]{}%
\providecommand \bibitemNoStop [0]{.\EOS\space}%
\providecommand \EOS [0]{\spacefactor3000\relax}%
\providecommand \BibitemShut  [1]{\csname bibitem#1\endcsname}%
\let\auto@bib@innerbib\@empty
%</preamble>
\bibitem [{\citenamefont {Horodecki}\ \emph {et~al.}(2009)\citenamefont
  {Horodecki}, \citenamefont {Horodecki}, \citenamefont {Horodecki},\ and\
  \citenamefont {Horodecki}}]{QE_RMP}%
  \BibitemOpen
  \bibfield  {author} {\bibinfo {author} {\bibfnamefont {R.}~\bibnamefont
  {Horodecki}}, \bibinfo {author} {\bibfnamefont {P.}~\bibnamefont
  {Horodecki}}, \bibinfo {author} {\bibfnamefont {M.}~\bibnamefont
  {Horodecki}}, \ and\ \bibinfo {author} {\bibfnamefont {K.}~\bibnamefont
  {Horodecki}},\ }\href {\doibase 10.1103/RevModPhys.81.865} {\bibfield
  {journal} {\bibinfo  {journal} {Rev. Mod. Phys.}\ }\textbf {\bibinfo {volume}
  {81}},\ \bibinfo {pages} {865} (\bibinfo {year} {2009})}\BibitemShut
  {NoStop}%
\bibitem [{\citenamefont {Eisert}\ \emph {et~al.}(2010)\citenamefont {Eisert},
  \citenamefont {Cramer},\ and\ \citenamefont {Plenio}}]{Eisert_RMP}%
  \BibitemOpen
  \bibfield  {author} {\bibinfo {author} {\bibfnamefont {J.}~\bibnamefont
  {Eisert}}, \bibinfo {author} {\bibfnamefont {M.}~\bibnamefont {Cramer}}, \
  and\ \bibinfo {author} {\bibfnamefont {M.~B.}\ \bibnamefont {Plenio}},\
  }\href {\doibase 10.1103/RevModPhys.82.277} {\bibfield  {journal} {\bibinfo
  {journal} {Rev. Mod. Phys.}\ }\textbf {\bibinfo {volume} {82}},\ \bibinfo
  {pages} {277} (\bibinfo {year} {2010})}\BibitemShut {NoStop}%
\bibitem [{\citenamefont {Calabrese}\ and\ \citenamefont
  {Cardy}(2004)}]{Calabrese2004}%
  \BibitemOpen
  \bibfield  {author} {\bibinfo {author} {\bibfnamefont {P.}~\bibnamefont
  {Calabrese}}\ and\ \bibinfo {author} {\bibfnamefont {J.}~\bibnamefont
  {Cardy}},\ }\href {http://stacks.iop.org/1742-5468/2004/i=06/a=P06002}
  {\bibfield  {journal} {\bibinfo  {journal} {Journal of Statistical Mechanics:
  Theory and Experiment}\ }\textbf {\bibinfo {volume} {2004}},\ \bibinfo
  {pages} {P06002} (\bibinfo {year} {2004})}\BibitemShut {NoStop}%
\bibitem [{\citenamefont {Refael}\ and\ \citenamefont
  {Moore}(2004)}]{Refael2004}%
  \BibitemOpen
  \bibfield  {author} {\bibinfo {author} {\bibfnamefont {G.}~\bibnamefont
  {Refael}}\ and\ \bibinfo {author} {\bibfnamefont {J.~E.}\ \bibnamefont
  {Moore}},\ }\href {\doibase 10.1103/PhysRevLett.93.260602} {\bibfield
  {journal} {\bibinfo  {journal} {Phys. Rev. Lett.}\ }\textbf {\bibinfo
  {volume} {93}},\ \bibinfo {pages} {260602} (\bibinfo {year}
  {2004})}\BibitemShut {NoStop}%
\bibitem [{\citenamefont {Laflorencie}(2005)}]{Laflorencie2005}%
  \BibitemOpen
  \bibfield  {author} {\bibinfo {author} {\bibfnamefont {N.}~\bibnamefont
  {Laflorencie}},\ }\href {\doibase 10.1103/PhysRevB.72.140408} {\bibfield
  {journal} {\bibinfo  {journal} {Phys. Rev. B}\ }\textbf {\bibinfo {volume}
  {72}},\ \bibinfo {pages} {140408} (\bibinfo {year} {2005})}\BibitemShut
  {NoStop}%
\bibitem [{\citenamefont {Bonesteel}\ and\ \citenamefont
  {Yang}(2007)}]{Bonesteel2007}%
  \BibitemOpen
  \bibfield  {author} {\bibinfo {author} {\bibfnamefont {N.~E.}\ \bibnamefont
  {Bonesteel}}\ and\ \bibinfo {author} {\bibfnamefont {K.}~\bibnamefont
  {Yang}},\ }\href {\doibase 10.1103/PhysRevLett.99.140405} {\bibfield
  {journal} {\bibinfo  {journal} {Phys. Rev. Lett.}\ }\textbf {\bibinfo
  {volume} {99}},\ \bibinfo {pages} {140405} (\bibinfo {year}
  {2007})}\BibitemShut {NoStop}%
\bibitem [{\citenamefont {Bravyi}\ \emph {et~al.}(2012)\citenamefont {Bravyi},
  \citenamefont {Caha}, \citenamefont {Movassagh}, \citenamefont {Nagaj},\ and\
  \citenamefont {Shor}}]{Bravyi2012}%
  \BibitemOpen
  \bibfield  {author} {\bibinfo {author} {\bibfnamefont {S.}~\bibnamefont
  {Bravyi}}, \bibinfo {author} {\bibfnamefont {L.}~\bibnamefont {Caha}},
  \bibinfo {author} {\bibfnamefont {R.}~\bibnamefont {Movassagh}}, \bibinfo
  {author} {\bibfnamefont {D.}~\bibnamefont {Nagaj}}, \ and\ \bibinfo {author}
  {\bibfnamefont {P.~W.}\ \bibnamefont {Shor}},\ }\href {\doibase
  10.1103/PhysRevLett.109.207202} {\bibfield  {journal} {\bibinfo  {journal}
  {Phys. Rev. Lett.}\ }\textbf {\bibinfo {volume} {109}},\ \bibinfo {pages}
  {207202} (\bibinfo {year} {2012})}\BibitemShut {NoStop}%
\bibitem [{\citenamefont {Wolf}(2006)}]{Wolf}%
  \BibitemOpen
  \bibfield  {author} {\bibinfo {author} {\bibfnamefont {M.~M.}\ \bibnamefont
  {Wolf}},\ }\href {\doibase 10.1103/PhysRevLett.96.010404} {\bibfield
  {journal} {\bibinfo  {journal} {Phys. Rev. Lett.}\ }\textbf {\bibinfo
  {volume} {96}},\ \bibinfo {pages} {010404} (\bibinfo {year}
  {2006})}\BibitemShut {NoStop}%
\bibitem [{\citenamefont {Gioev}\ and\ \citenamefont
  {Klich}(2006)}]{GioevKlich}%
  \BibitemOpen
  \bibfield  {author} {\bibinfo {author} {\bibfnamefont {D.}~\bibnamefont
  {Gioev}}\ and\ \bibinfo {author} {\bibfnamefont {I.}~\bibnamefont {Klich}},\
  }\href {\doibase 10.1103/PhysRevLett.96.100503} {\bibfield  {journal}
  {\bibinfo  {journal} {Phys. Rev. Lett.}\ }\textbf {\bibinfo {volume} {96}},\
  \bibinfo {pages} {100503} (\bibinfo {year} {2006})}\BibitemShut {NoStop}%
\bibitem [{\citenamefont {Swingle}(2010)}]{Swingle}%
  \BibitemOpen
  \bibfield  {author} {\bibinfo {author} {\bibfnamefont {B.}~\bibnamefont
  {Swingle}},\ }\href {\doibase 10.1103/PhysRevLett.105.050502} {\bibfield
  {journal} {\bibinfo  {journal} {Phys. Rev. Lett.}\ }\textbf {\bibinfo
  {volume} {105}},\ \bibinfo {pages} {050502} (\bibinfo {year}
  {2010})}\BibitemShut {NoStop}%
\bibitem [{\citenamefont {Ding}\ \emph {et~al.}(2012)\citenamefont {Ding},
  \citenamefont {Seidel},\ and\ \citenamefont {Yang}}]{dingprx12}%
  \BibitemOpen
  \bibfield  {author} {\bibinfo {author} {\bibfnamefont {W.}~\bibnamefont
  {Ding}}, \bibinfo {author} {\bibfnamefont {A.}~\bibnamefont {Seidel}}, \ and\
  \bibinfo {author} {\bibfnamefont {K.}~\bibnamefont {Yang}},\ }\href {\doibase
  10.1103/PhysRevX.2.011012} {\bibfield  {journal} {\bibinfo  {journal} {Phys.
  Rev. X}\ }\textbf {\bibinfo {volume} {2}},\ \bibinfo {pages} {011012}
  (\bibinfo {year} {2012})}\BibitemShut {NoStop}%
\bibitem [{\citenamefont {Lai}\ \emph {et~al.}(2013)\citenamefont {Lai},
  \citenamefont {Yang},\ and\ \citenamefont {Bonesteel}}]{Lai_EE4EBL}%
  \BibitemOpen
  \bibfield  {author} {\bibinfo {author} {\bibfnamefont {H.-H.}\ \bibnamefont
  {Lai}}, \bibinfo {author} {\bibfnamefont {K.}~\bibnamefont {Yang}}, \ and\
  \bibinfo {author} {\bibfnamefont {N.~E.}\ \bibnamefont {Bonesteel}},\ }\href
  {\doibase 10.1103/PhysRevLett.111.210402} {\bibfield  {journal} {\bibinfo
  {journal} {Phys. Rev. Lett.}\ }\textbf {\bibinfo {volume} {111}},\ \bibinfo
  {pages} {210402} (\bibinfo {year} {2013})}\BibitemShut {NoStop}%
\bibitem [{\citenamefont {Paramekanti}\ \emph {et~al.}(2002)\citenamefont
  {Paramekanti}, \citenamefont {Balents},\ and\ \citenamefont
  {Fisher}}]{Paramekanti02}%
  \BibitemOpen
  \bibfield  {author} {\bibinfo {author} {\bibfnamefont {A.}~\bibnamefont
  {Paramekanti}}, \bibinfo {author} {\bibfnamefont {L.}~\bibnamefont
  {Balents}}, \ and\ \bibinfo {author} {\bibfnamefont {M.~P.~A.}\ \bibnamefont
  {Fisher}},\ }\href {\doibase 10.1103/PhysRevB.66.054526} {\bibfield
  {journal} {\bibinfo  {journal} {Phys. Rev. B}\ }\textbf {\bibinfo {volume}
  {66}},\ \bibinfo {pages} {054526} (\bibinfo {year} {2002})}\BibitemShut
  {NoStop}%
\bibitem [{\citenamefont {Tay}\ and\ \citenamefont {Motrunich}(2010)}]{Tay10}%
  \BibitemOpen
  \bibfield  {author} {\bibinfo {author} {\bibfnamefont {T.}~\bibnamefont
  {Tay}}\ and\ \bibinfo {author} {\bibfnamefont {O.~I.}\ \bibnamefont
  {Motrunich}},\ }\href {\doibase 10.1103/PhysRevLett.105.187202} {\bibfield
  {journal} {\bibinfo  {journal} {Phys. Rev. Lett.}\ }\textbf {\bibinfo
  {volume} {105}},\ \bibinfo {pages} {187202} (\bibinfo {year}
  {2010})}\BibitemShut {NoStop}%
\bibitem [{\citenamefont {Zhang}\ \emph {et~al.}(2011)\citenamefont {Zhang},
  \citenamefont {Grover},\ and\ \citenamefont {Vishwanath}}]{Zhang11}%
  \BibitemOpen
  \bibfield  {author} {\bibinfo {author} {\bibfnamefont {Y.}~\bibnamefont
  {Zhang}}, \bibinfo {author} {\bibfnamefont {T.}~\bibnamefont {Grover}}, \
  and\ \bibinfo {author} {\bibfnamefont {A.}~\bibnamefont {Vishwanath}},\
  }\href@noop {} {\bibfield  {journal} {\bibinfo  {journal} {Phys. Rev. Lett.}\
  }\textbf {\bibinfo {volume} {107}},\ \bibinfo {pages} {067202} (\bibinfo
  {year} {2011})}\BibitemShut {NoStop}%
\bibitem [{\citenamefont {Shao}\ \emph {et~al.}(2015)\citenamefont {Shao},
  \citenamefont {Kim}, \citenamefont {Haldane},\ and\ \citenamefont
  {Rezayi}}]{EE_HLR}%
  \BibitemOpen
  \bibfield  {author} {\bibinfo {author} {\bibfnamefont {J.}~\bibnamefont
  {Shao}}, \bibinfo {author} {\bibfnamefont {E.-A.}\ \bibnamefont {Kim}},
  \bibinfo {author} {\bibfnamefont {F.~D.~M.}\ \bibnamefont {Haldane}}, \ and\
  \bibinfo {author} {\bibfnamefont {E.~H.}\ \bibnamefont {Rezayi}},\ }\href
  {\doibase 10.1103/PhysRevLett.114.206402} {\bibfield  {journal} {\bibinfo
  {journal} {Phys. Rev. Lett.}\ }\textbf {\bibinfo {volume} {114}},\ \bibinfo
  {pages} {206402} (\bibinfo {year} {2015})}\BibitemShut {NoStop}%
\bibitem [{\citenamefont {Ryu}\ and\ \citenamefont
  {Takayanagi}(2006)}]{RyuTakayanagi2006}%
  \BibitemOpen
  \bibfield  {author} {\bibinfo {author} {\bibfnamefont {S.}~\bibnamefont
  {Ryu}}\ and\ \bibinfo {author} {\bibfnamefont {T.}~\bibnamefont
  {Takayanagi}},\ }\href {http://stacks.iop.org/1126-6708/2006/i=08/a=045}
  {\bibfield  {journal} {\bibinfo  {journal} {Journal of High Energy Physics}\
  }\textbf {\bibinfo {volume} {2006}},\ \bibinfo {pages} {045} (\bibinfo {year}
  {2006})}\BibitemShut {NoStop}%
\bibitem [{\citenamefont {Nishioka}\ \emph {et~al.}(2009)\citenamefont
  {Nishioka}, \citenamefont {Ryu},\ and\ \citenamefont
  {Takayanagi}}]{Tatsuma2009}%
  \BibitemOpen
  \bibfield  {author} {\bibinfo {author} {\bibfnamefont {T.}~\bibnamefont
  {Nishioka}}, \bibinfo {author} {\bibfnamefont {S.}~\bibnamefont {Ryu}}, \
  and\ \bibinfo {author} {\bibfnamefont {T.}~\bibnamefont {Takayanagi}},\
  }\href {http://stacks.iop.org/1751-8121/42/i=50/a=504008} {\bibfield
  {journal} {\bibinfo  {journal} {Journal of Physics A: Mathematical and
  Theoretical}\ }\textbf {\bibinfo {volume} {42}},\ \bibinfo {pages} {504008}
  (\bibinfo {year} {2009})}\BibitemShut {NoStop}%
\bibitem [{\citenamefont {Ogawa}\ \emph {et~al.}(2012)\citenamefont {Ogawa},
  \citenamefont {Takayanagi},\ and\ \citenamefont {Ugajin}}]{Ogawa2012}%
  \BibitemOpen
  \bibfield  {author} {\bibinfo {author} {\bibfnamefont {N.}~\bibnamefont
  {Ogawa}}, \bibinfo {author} {\bibfnamefont {T.}~\bibnamefont {Takayanagi}}, \
  and\ \bibinfo {author} {\bibfnamefont {T.}~\bibnamefont {Ugajin}},\ }\href
  {http://dx.doi.org/10.1007/JHEP01} {\bibfield  {journal} {\bibinfo  {journal}
  {Journal of High Energy Physics}\ }\textbf {\bibinfo {volume} {2012}},\
  \bibinfo {eid} {125} (\bibinfo {year} {2012})}\BibitemShut {NoStop}%
\bibitem [{\citenamefont {Huijse}\ \emph {et~al.}(2012)\citenamefont {Huijse},
  \citenamefont {Sachdev},\ and\ \citenamefont {Swingle}}]{Huijse12}%
  \BibitemOpen
  \bibfield  {author} {\bibinfo {author} {\bibfnamefont {L.}~\bibnamefont
  {Huijse}}, \bibinfo {author} {\bibfnamefont {S.}~\bibnamefont {Sachdev}}, \
  and\ \bibinfo {author} {\bibfnamefont {B.}~\bibnamefont {Swingle}},\ }\href
  {\doibase 10.1103/PhysRevB.85.035121} {\bibfield  {journal} {\bibinfo
  {journal} {Phys. Rev. B}\ }\textbf {\bibinfo {volume} {85}},\ \bibinfo
  {pages} {035121} (\bibinfo {year} {2012})}\BibitemShut {NoStop}%
\bibitem [{\citenamefont {Shoenberg}(1984)}]{Book-Shoenberg}%
  \BibitemOpen
  \bibfield  {author} {\bibinfo {author} {\bibfnamefont {D.~D.}\ \bibnamefont
  {Shoenberg}},\ }\href@noop {} {\emph {\bibinfo {title} {Magnetic oscillations
  in metals}}}\ (\bibinfo  {publisher} {Cambridge University Press},\ \bibinfo
  {year} {1984})\BibitemShut {NoStop}%
\bibitem [{\citenamefont {Cramer}\ \emph {et~al.}(2007)\citenamefont {Cramer},
  \citenamefont {Eisert},\ and\ \citenamefont {Plenio}}]{cramer2007statistics}%
  \BibitemOpen
  \bibfield  {author} {\bibinfo {author} {\bibfnamefont {M.}~\bibnamefont
  {Cramer}}, \bibinfo {author} {\bibfnamefont {J.}~\bibnamefont {Eisert}}, \
  and\ \bibinfo {author} {\bibfnamefont {M.~B.}\ \bibnamefont {Plenio}},\
  }\href {\doibase 10.1103/PhysRevLett.98.220603} {\bibfield  {journal}
  {\bibinfo  {journal} {Phys. Rev. Lett.}\ }\textbf {\bibinfo {volume} {98}},\
  \bibinfo {pages} {220603} (\bibinfo {year} {2007})}\BibitemShut {NoStop}%
\bibitem [{Inv()}]{Inversion_center}%
  \BibitemOpen
  \href@noop {} {}\bibinfo {note} {The assumption is quite natural for any
  system with Fermi surface preserving time-reversal symmetry. For a system
  with Bose surface, we are not certain how the low-energy degrees of freedom
  transform under time-reversal symmetry, so we may require an inversion
  symmetry or $C_2$ rotation symmetry, which is indeed preserved in the only
  well-established system with a Bose surface--EBL \cite{Paramekanti02,Tay10,
  Lai_EE4EBL}.}\BibitemShut {Stop}%
\bibitem [{Rad()}]{Radon_Transform}%
  \BibitemOpen
  \href@noop {} {}\bibinfo {note} {Our method in 2D is \textit{different} to
  the widely used technique of shape reconstruction known as Radon transform
  \cite{Book-Radon} in tomography, the creation of an image from the projection
  data associated with cross-sectional scans of an object. The Radon function
  computes projections of an object image matrix along specified directions and
  it is a set of \textit{line integral} along 1D arrays perpendicular to the
  axis where the object is projected onto. The Radon function at a specified
  direction, say $\hat{x}$, contains the information of \textit{both} the
  projected length along $\hat{x}$ and the heights as a function of $x$ along
  the vertical axis perpendicular $\hat{x}$. However, our EE probe can only
  give the information of projected lengths at specified
  directions.}\BibitemShut {Stop}%
\bibitem [{\citenamefont {Luttinger}(1960)}]{Luttinger1960}%
  \BibitemOpen
  \bibfield  {author} {\bibinfo {author} {\bibfnamefont {J.~M.}\ \bibnamefont
  {Luttinger}},\ }\href {\doibase 10.1103/PhysRev.119.1153} {\bibfield
  {journal} {\bibinfo  {journal} {Phys. Rev.}\ }\textbf {\bibinfo {volume}
  {119}},\ \bibinfo {pages} {1153} (\bibinfo {year} {1960})}\BibitemShut
  {NoStop}%
\bibitem [{\citenamefont {Friedel}(1954)}]{Friedel}%
  \BibitemOpen
  \bibfield  {author} {\bibinfo {author} {\bibfnamefont {J.}~\bibnamefont
  {Friedel}},\ }\href@noop {} {\bibfield  {journal} {\bibinfo  {journal}
  {Advances in Physics}\ }\textbf {\bibinfo {volume} {3}},\ \bibinfo {pages}
  {446} (\bibinfo {year} {1954})}\BibitemShut {NoStop}%
\bibitem [{con()}]{concave_assumption}%
  \BibitemOpen
  \href@noop {} {}\bibinfo {note} {We assume that the concavity is large enough
  so that number of additional intersected points contribute enough leading EE
  to make the line upward; if the number of additional intersected points is
  not big enough, the slope may not change sign but the cusp is still
  there}\BibitemShut {NoStop}%
\bibitem [{\citenamefont {Lee}\ and\ \citenamefont {Lee}(2005)}]{LeeandLee05}%
  \BibitemOpen
  \bibfield  {author} {\bibinfo {author} {\bibfnamefont {S.-S.}\ \bibnamefont
  {Lee}}\ and\ \bibinfo {author} {\bibfnamefont {P.~A.}\ \bibnamefont {Lee}},\
  }\href {\doibase 10.1103/PhysRevLett.95.036403} {\bibfield  {journal}
  {\bibinfo  {journal} {Phys. Rev. Lett.}\ }\textbf {\bibinfo {volume} {95}},\
  \bibinfo {pages} {036403} (\bibinfo {year} {2005})}\BibitemShut {NoStop}%
\bibitem [{\citenamefont {Motrunich}(2005)}]{ringxch}%
  \BibitemOpen
  \bibfield  {author} {\bibinfo {author} {\bibfnamefont {O.~I.}\ \bibnamefont
  {Motrunich}},\ }\href@noop {} {\bibfield  {journal} {\bibinfo  {journal}
  {Phys. Rev. B}\ }\textbf {\bibinfo {volume} {72}},\ \bibinfo {pages} {045105}
  (\bibinfo {year} {2005})}\BibitemShut {NoStop}%
\bibitem [{\citenamefont {Sheng}\ \emph {et~al.}(2009)\citenamefont {Sheng},
  \citenamefont {Motrunich},\ and\ \citenamefont {Fisher}}]{Sheng09}%
  \BibitemOpen
  \bibfield  {author} {\bibinfo {author} {\bibfnamefont {D.~N.}\ \bibnamefont
  {Sheng}}, \bibinfo {author} {\bibfnamefont {O.~I.}\ \bibnamefont
  {Motrunich}}, \ and\ \bibinfo {author} {\bibfnamefont {M.~P.~A.}\
  \bibnamefont {Fisher}},\ }\href@noop {} {\bibfield  {journal} {\bibinfo
  {journal} {Phys. Rev. B}\ }\textbf {\bibinfo {volume} {79}},\ \bibinfo
  {pages} {205112} (\bibinfo {year} {2009})}\BibitemShut {NoStop}%
\bibitem [{\citenamefont {Block}\ \emph {et~al.}(2011)\citenamefont {Block},
  \citenamefont {Sheng}, \citenamefont {Motrunich},\ and\ \citenamefont
  {Fisher}}]{Block2010}%
  \BibitemOpen
  \bibfield  {author} {\bibinfo {author} {\bibfnamefont {M.~S.}\ \bibnamefont
  {Block}}, \bibinfo {author} {\bibfnamefont {D.~N.}\ \bibnamefont {Sheng}},
  \bibinfo {author} {\bibfnamefont {O.~I.}\ \bibnamefont {Motrunich}}, \ and\
  \bibinfo {author} {\bibfnamefont {M.~P.~A.}\ \bibnamefont {Fisher}},\ }\href
  {\doibase 10.1103/PhysRevLett.106.157202} {\bibfield  {journal} {\bibinfo
  {journal} {Phys. Rev. Lett.}\ }\textbf {\bibinfo {volume} {106}},\ \bibinfo
  {pages} {157202} (\bibinfo {year} {2011})}\BibitemShut {NoStop}%
\bibitem [{\citenamefont {Mishmash}\ \emph {et~al.}(2011)\citenamefont
  {Mishmash}, \citenamefont {Block}, \citenamefont {Kaul}, \citenamefont
  {Sheng}, \citenamefont {Motrunich},\ and\ \citenamefont
  {Fisher}}]{Mishmash11}%
  \BibitemOpen
  \bibfield  {author} {\bibinfo {author} {\bibfnamefont {R.~V.}\ \bibnamefont
  {Mishmash}}, \bibinfo {author} {\bibfnamefont {M.~S.}\ \bibnamefont {Block}},
  \bibinfo {author} {\bibfnamefont {R.~K.}\ \bibnamefont {Kaul}}, \bibinfo
  {author} {\bibfnamefont {D.~N.}\ \bibnamefont {Sheng}}, \bibinfo {author}
  {\bibfnamefont {O.~I.}\ \bibnamefont {Motrunich}}, \ and\ \bibinfo {author}
  {\bibfnamefont {M.~P.~A.}\ \bibnamefont {Fisher}},\ }\href {\doibase
  10.1103/PhysRevB.84.245127} {\bibfield  {journal} {\bibinfo  {journal} {Phys.
  Rev. B}\ }\textbf {\bibinfo {volume} {84}},\ \bibinfo {pages} {245127}
  (\bibinfo {year} {2011})}\BibitemShut {NoStop}%
\bibitem [{\citenamefont {Chua}\ \emph {et~al.}(2011)\citenamefont {Chua},
  \citenamefont {Yao},\ and\ \citenamefont {Fiete}}]{Chua11}%
  \BibitemOpen
  \bibfield  {author} {\bibinfo {author} {\bibfnamefont {V.}~\bibnamefont
  {Chua}}, \bibinfo {author} {\bibfnamefont {H.}~\bibnamefont {Yao}}, \ and\
  \bibinfo {author} {\bibfnamefont {G.~A.}\ \bibnamefont {Fiete}},\ }\href
  {\doibase 10.1103/PhysRevB.83.180412} {\bibfield  {journal} {\bibinfo
  {journal} {Phys. Rev. B}\ }\textbf {\bibinfo {volume} {83}},\ \bibinfo
  {pages} {180412} (\bibinfo {year} {2011})}\BibitemShut {NoStop}%
\bibitem [{\citenamefont {Biswas}\ \emph {et~al.}(2011)\citenamefont {Biswas},
  \citenamefont {Fu}, \citenamefont {Laumann},\ and\ \citenamefont
  {Sachdev}}]{Biswas11}%
  \BibitemOpen
  \bibfield  {author} {\bibinfo {author} {\bibfnamefont {R.~R.}\ \bibnamefont
  {Biswas}}, \bibinfo {author} {\bibfnamefont {L.}~\bibnamefont {Fu}}, \bibinfo
  {author} {\bibfnamefont {C.~R.}\ \bibnamefont {Laumann}}, \ and\ \bibinfo
  {author} {\bibfnamefont {S.}~\bibnamefont {Sachdev}},\ }\href {\doibase
  10.1103/PhysRevB.83.245131} {\bibfield  {journal} {\bibinfo  {journal} {Phys.
  Rev. B}\ }\textbf {\bibinfo {volume} {83}},\ \bibinfo {pages} {245131}
  (\bibinfo {year} {2011})}\BibitemShut {NoStop}%
\bibitem [{\citenamefont {Baskaran}\ \emph {et~al.}(shed)\citenamefont
  {Baskaran}, \citenamefont {Santhosh},\ and\ \citenamefont
  {Shankar}}]{Baskaran09}%
  \BibitemOpen
  \bibfield  {author} {\bibinfo {author} {\bibfnamefont {G.}~\bibnamefont
  {Baskaran}}, \bibinfo {author} {\bibfnamefont {G.}~\bibnamefont {Santhosh}},
  \ and\ \bibinfo {author} {\bibfnamefont {R.}~\bibnamefont {Shankar}},\
  }\href@noop {} {\bibfield  {journal} {\bibinfo  {journal}
  {arXiv:0908.1614v3}\ } (\bibinfo {year} {unpublished})}\BibitemShut {NoStop}%
\bibitem [{\citenamefont {Lai}\ and\ \citenamefont {Motrunich}(2011)}]{Lai11}%
  \BibitemOpen
  \bibfield  {author} {\bibinfo {author} {\bibfnamefont {H.-H.}\ \bibnamefont
  {Lai}}\ and\ \bibinfo {author} {\bibfnamefont {O.~I.}\ \bibnamefont
  {Motrunich}},\ }\href {\doibase 10.1103/PhysRevB.84.085141} {\bibfield
  {journal} {\bibinfo  {journal} {Phys. Rev. B}\ }\textbf {\bibinfo {volume}
  {84}},\ \bibinfo {pages} {085141} (\bibinfo {year} {2011})}\BibitemShut
  {NoStop}%
\bibitem [{\citenamefont {Hermanns}\ and\ \citenamefont
  {Trebst}(2014)}]{Hermanns2014}%
  \BibitemOpen
  \bibfield  {author} {\bibinfo {author} {\bibfnamefont {M.}~\bibnamefont
  {Hermanns}}\ and\ \bibinfo {author} {\bibfnamefont {S.}~\bibnamefont
  {Trebst}},\ }\href {\doibase 10.1103/PhysRevB.89.235102} {\bibfield
  {journal} {\bibinfo  {journal} {Phys. Rev. B}\ }\textbf {\bibinfo {volume}
  {89}},\ \bibinfo {pages} {235102} (\bibinfo {year} {2014})}\BibitemShut
  {NoStop}%
\bibitem [{\citenamefont {Lai}\ and\ \citenamefont {Motrunich}(2010)}]{Lai10}%
  \BibitemOpen
  \bibfield  {author} {\bibinfo {author} {\bibfnamefont {H.-H.}\ \bibnamefont
  {Lai}}\ and\ \bibinfo {author} {\bibfnamefont {O.~I.}\ \bibnamefont
  {Motrunich}},\ }\href@noop {} {\bibfield  {journal} {\bibinfo  {journal}
  {Phys. Rev. B}\ }\textbf {\bibinfo {volume} {81}},\ \bibinfo {pages} {045105}
  (\bibinfo {year} {2010})}\BibitemShut {NoStop}%
\bibitem [{\citenamefont {Mishmash}\ \emph {et~al.}(2015)\citenamefont
  {Mishmash}, \citenamefont {Gonz\'alez}, \citenamefont {Melko}, \citenamefont
  {Motrunich},\ and\ \citenamefont {Fisher}}]{Mishmash2015}%
  \BibitemOpen
  \bibfield  {author} {\bibinfo {author} {\bibfnamefont {R.~V.}\ \bibnamefont
  {Mishmash}}, \bibinfo {author} {\bibfnamefont {I.}~\bibnamefont
  {Gonz\'alez}}, \bibinfo {author} {\bibfnamefont {R.~G.}\ \bibnamefont
  {Melko}}, \bibinfo {author} {\bibfnamefont {O.~I.}\ \bibnamefont
  {Motrunich}}, \ and\ \bibinfo {author} {\bibfnamefont {M.~P.~A.}\
  \bibnamefont {Fisher}},\ }\href {\doibase 10.1103/PhysRevB.91.235140}
  {\bibfield  {journal} {\bibinfo  {journal} {Phys. Rev. B}\ }\textbf {\bibinfo
  {volume} {91}},\ \bibinfo {pages} {235140} (\bibinfo {year}
  {2015})}\BibitemShut {NoStop}%
\bibitem [{\citenamefont {Motrunich}\ and\ \citenamefont {Fisher}(2007)}]{DBL}%
  \BibitemOpen
  \bibfield  {author} {\bibinfo {author} {\bibfnamefont {O.~I.}\ \bibnamefont
  {Motrunich}}\ and\ \bibinfo {author} {\bibfnamefont {M.~P.~A.}\ \bibnamefont
  {Fisher}},\ }\href@noop {} {\bibfield  {journal} {\bibinfo  {journal} {Phys.
  Rev. B}\ }\textbf {\bibinfo {volume} {75}},\ \bibinfo {pages} {235116}
  (\bibinfo {year} {2007})}\BibitemShut {NoStop}%
\bibitem [{\citenamefont {Jiang}\ \emph {et~al.}(2013)\citenamefont {Jiang},
  \citenamefont {Block}, \citenamefont {Mishmash}, \citenamefont {Garrison},
  \citenamefont {Sheng}, \citenamefont {Motrunich},\ and\ \citenamefont
  {Fisher}}]{Jiang2013}%
  \BibitemOpen
  \bibfield  {author} {\bibinfo {author} {\bibfnamefont {H.-C.}\ \bibnamefont
  {Jiang}}, \bibinfo {author} {\bibfnamefont {M.~S.}\ \bibnamefont {Block}},
  \bibinfo {author} {\bibfnamefont {R.~V.}\ \bibnamefont {Mishmash}}, \bibinfo
  {author} {\bibfnamefont {J.~R.}\ \bibnamefont {Garrison}}, \bibinfo {author}
  {\bibfnamefont {D.~N.}\ \bibnamefont {Sheng}}, \bibinfo {author}
  {\bibfnamefont {O.~I.}\ \bibnamefont {Motrunich}}, \ and\ \bibinfo {author}
  {\bibfnamefont {M.~P.~A.}\ \bibnamefont {Fisher}},\ }\href {\doibase
  10.1038/nature11732} {\bibfield  {journal} {\bibinfo  {journal} {Nature}\
  }\textbf {\bibinfo {volume} {493}},\ \bibinfo {pages} {39} (\bibinfo {year}
  {2013})}\BibitemShut {NoStop}%
\bibitem [{\citenamefont {Lee}\ \emph {et~al.}(2006)\citenamefont {Lee},
  \citenamefont {Nagaosa},\ and\ \citenamefont {Wen}}]{LeeNagaosaWen}%
  \BibitemOpen
  \bibfield  {author} {\bibinfo {author} {\bibfnamefont {P.~A.}\ \bibnamefont
  {Lee}}, \bibinfo {author} {\bibfnamefont {N.}~\bibnamefont {Nagaosa}}, \ and\
  \bibinfo {author} {\bibfnamefont {X.-G.}\ \bibnamefont {Wen}},\ }\href@noop
  {} {\bibfield  {journal} {\bibinfo  {journal} {Rev. Mod. Phys.}\ }\textbf
  {\bibinfo {volume} {78}},\ \bibinfo {pages} {17} (\bibinfo {year}
  {2006})}\BibitemShut {NoStop}%
\bibitem [{\citenamefont {Deans}(2007)}]{Book-Radon}%
  \BibitemOpen
  \bibfield  {author} {\bibinfo {author} {\bibfnamefont {S.~R.}\ \bibnamefont
  {Deans}},\ }\href@noop {} {\emph {\bibinfo {title} {The Radon Transform and
  Some of Its Applications}}}\ (\bibinfo  {publisher} {Dover Publications,
  Inc},\ \bibinfo {year} {2007})\BibitemShut {NoStop}%
\end{thebibliography}%
\end{document}